\documentclass{article}

\newcommand{\Ket}[1]{\left\vert#1\right\rangle}

\newcommand{\BraKet}[2]{\left\langle#1\right\vert\left.#2\right\rangle}
\newcommand{\KetBra}[2]{\left\vert#1\right\rangle\left\langle#2\right\vert}

\newcommand{\MatrixEl}[3]{\left\langle#1\right\vert #2 \left\vert#3\right\rangle}

\newcommand{\vett}[1]{\stackrel{\rightarrow}{#1}}

\newcommand{\NumOpWithLow}[2]{\hat{#1}^\dag_#2\hat{#1}_#2}
\newcommand{\NumSum}[1]{\sum_{j=x,y,z}\NumOpWithLow{#1}{j}}

\title{Realization of a Space Reversal Operator}

\author{B.Militello, A. Messina, A. Napoli \\INFM, MURST\\ and Dipartimento di Scienze Fisiche ed Astronomiche \\Via archirafi 36, 90123 Palermo (ITALY) \\ e-mail: messina@fisica.unipa.it}
\begin{document}

\maketitle

\begin{abstract}
       In this paper we propose the realization of a bosonic-fermionic interaction in the context of trapped ions whose effect upon the ion center of mass degrees of freedom is properly speaking a spatial inversion. The physical system and its features are accurately described and some applications are briefly discussed.
\end{abstract}

\section{Introduction}

The individuation of the symmetries of a given physical system plays a role of fundamental importance to understand its dynamical behavior. Each symmetry operation may be associated to a unitary operator depending on a continuous parameter or having a discrete nature. Rotations about a given axis or translations along prefixed directions provide examples of continuous symmetry transformations. Reflections with respect to planes are well known examples of discrete symmetry operators.

In this paper we focus our attention on the space reversion and specular reflections operators. Although the action of such operators can be mathematically simply described, their implementation in laboratory, at the best of our knowledge, has never been proposed up to now. 
The main result of this paper is the presentation of a method to realize these operators in the context of trapped ions. 

The ion confinement may be for example realized using a Paul micro trap where a suitable configuration of inhomogeneous and alternating electromagnetic fields forces the ion centre of mass to behave as a quantum harmonic oscillator\cite{Ghosh,Toschek}.
When in addition appropriate laser beams act upon the trapped ion, a coupling between ionic electronic and vibrational degrees of freedom is switched on\cite{nist, Vogel-Rass, VogelModelli}.
Such an interaction, sometimes called vibronic, allows the controlled manipulation of the trapped ion quantum state with possible applications in quantum logic context\cite{logic} and in the generation of non-classical behaviours\cite{Vario}.

Our scheme is based on the introduction and implementation of an appropriate vibronic coupling.
Under the action of this interaction the system, prepared in its fermionic ground state and without any prescription on the vibrational part, evolves toward a fermionic-bosonic disentangled target state reached after a well controlled interval of time. The vibrational part of such a target state coincides with that one obtains applying for example the space reversal or a specular reflection operator to the initial vibrational state. In this sense the time evolution under the prescribed vibronic coupling is said to realize the operator under scrutiny.
Generally speaking, in the last years trapped ions have been largely used to investigate fundamental aspects of quantum mechanics\cite{nist,VogelModelli,Vario,nist-exp}. This is due to the possibility of implementing a huge class of vibronic couplings\cite{nist,VogelModelli,Knight} and to the absence of decoherence during time intervals of experimental interest wide enough to observe \emph{pure} quantum mechanical phenomena\cite{nist,VogelModelli,nist-exp}.
Apart from its intrinsic theoretical interest, the realization of a space reversal operator or of a specular reflection operator, provides a good starting point for some attractive applications. In the last section we indeed show how our method can be exploited to generate a state orthogonal to a given one and to reverse a huge class of time evolutions. Such a reversal means restoring the state of the system at $t_0$ from the state reached at a generic time instant $t$, given that $t_0<t$.
The paper is organized as follows. In the next section we introduce our notation and recall definition and properties of the operators of interest.
In Section 3 we describe the physical system and a class of vibronic couplings allowing the realization of the scheme whose details are reported in section 4. Some applications and conclusive comments are finally discussed in the last two sections.

\section{Specular reflection and space reversal operators}

Take into account the state $\Ket{\vett{r}_1,\vett{r}_2,...,\vett{r}_n}$ which corresponds to the $n$ localized particles belonging to the system under scrutiny.
The action of the operator ($\hat{\Pi}$) describing a space inversion may be mathematically written down in the following manner
\begin{equation}\label{equivalence} 
  \hat{\Pi}\Ket{\vett{r}_1,\vett{r}_2,...,\vett{r}_n} = \Ket{-\vett{r}_1,-\vett{r}_2,...,-\vett{r}_n}
\end{equation}\\

In the same way it may be written down the action of the operator describing a specular reflection with respect to a plane or, in other words, describing the reversion of an axis. Let $z$ be such an axis and $xy$ the orthogonal plane. Calling $\hat{\Pi}_z$ the related specular reflection operator we have
\begin{equation}\label{equivalence} 
  \hat{\Pi}_z\Ket{\vett{r}_1,\vett{r}_2,...,\vett{r}_n} =              \Ket{\vett{r}_1^*,\vett{r}_2^*,...,\vett{r}_n^*}
\end{equation}\\
where $\vett{r}_i^*\equiv (x_i,y_i,-z_i)$ if $\vett{r}_i\equiv (x_i,y_i,z_i)$.

From these definitions it immediately follows that the effect of a space reversal operator ($\hat{\Pi}$) is equivalent to the action of the three operators ($\Pi_i$, with $i=x,y,z$) describing three specular reflections with respect to three orthogonal planes ($yz$,$xz$,$xy$).
\begin{equation}\label{equivalence} 
  \hat{\Pi} = \hat{\Pi}_x \hat{\Pi}_y \hat{\Pi}_z
\end{equation}\\

Hence the target of implementing both space reversal and specular reflection operators may be reached just engineering the second ones. This is the purpose of the following sections.

\section{The physical system}

Consider a single two effective level ion trapped in a three-dimensional isotropic Paul trap\cite{Ghosh,Toschek}. In such a situation the motion of the particle center of mass is harmonically confined via suitable inhomogeneous and alternating electromagnetic fields so that the ion free Hamiltonian is  
\begin{eqnarray}
  \hat{H}_0=\hbar{\omega}_1\KetBra{-}{-}+\hbar{\omega}_2\KetBra{+}{+}             
             +\hbar\omega_0\NumSum{a}
\end{eqnarray}

Here $\Ket{-}$ and $\Ket{+}$ indicate the ground and excited atomic (electronic) states and  $\omega_A=\omega_2-\omega_1$ is the related atomic transition frequency, while $\hat{a}_j$ ($\hat{a}_j^{\dag}$) with $j=x,y,z$ is the annihilation (creation) operator relative to the effective harmonic motion of the c.m. of the trapped ion (bosonic degrees of freedom).

Implementing a desired vibronic coupling is often the result of a Raman scheme, involving two \emph{virtual} transitions and an \emph{effective} one between $\Ket{+}$ and $\Ket{-}$. Using this scheme when the two levels are not dipolarly coupled allows to neglect the effects of spontaneous emissions\cite{Knight}.
Consider an auxiliary atomic level $\Ket{v}$ of eigenenergy $\hbar\omega_v$ and the action of two lasers of frequencies, wave vectors and amplitudes $\omega_{a}$, $\vett{k}_{a}$ and $\vett{E}_{a}$ ($\omega_{b}$, $\vett{k}_{b}$ and $\vett{E}_{b}$) respectively. Let the dipole transitions $\Ket{-}\longleftrightarrow\Ket{+}$ be forbidden, and the dipole transitions $\Ket{\pm}\longleftrightarrow\Ket{v}$ permitted. In addition suppose the frequency $\omega_{a}$ ($\omega_{b}$) be near the atomic transition frequency $\omega_v-\omega_1$ ($\omega_v-\omega_2$), but yet off-resonant.
The Schr\"odinger picture hamiltonian of the system, in the Rotating Wave Approximation (RWA), may be cast in the form
\begin{eqnarray}\label{Ham-Complete-2}
  \nonumber
  \hat{H}&=&\hat{H}_0+\hbar\omega_v\KetBra{v}{v}+\\
         &-&\left[\hbar g_{a} e^{-i(\vett{k}_{a}\cdot \vett{r}-\omega_{a}t)}\KetBra{-}{v}
         +\hbar g_{b} e^{-i(\vett{k}_{b}\cdot \vett{r}-\omega_{b}t)}\KetBra{+}{v} 
         + h.c. \right]
\end{eqnarray}
where $g_{a}=\MatrixEl{-}{\vett{d}\cdot\vett{E}_{a}}{v}$ ($g_{b}=\MatrixEl{+}{\vett{d}\cdot\vett{E}_{b}}{v}$), $\vett{d}$ being the atomic dipole operator.

In the spirit of the procedure followed in ref.\cite{Knight} we suppose that the large detunings and small coupling conditions are satisfied. Thus, putting
$|\Delta_{+v}|\equiv |\omega_b-(\omega_v-\omega_2)|$ and
$|\Delta_{-v}|\equiv |\omega_a-(\omega_v-\omega_1)|$, we have
\begin{equation}\label{Ham-Complete-2}
  |\Delta_{+v}|,|\Delta_{-v}|>> 
  |g_a|,|g_b|,|\Delta_{-v}-\Delta_{+v}|
\end{equation}

As a consequence of such an hypothesis one may assume that transitions $\Ket{\pm}\leftrightarrow\Ket{v}$ are too rare because of the \emph{non conservation} of energy in these processes\cite{virtual}. Ref\cite{virtual} clearly shows that an initially non populated atomic level, coupled to other two levels by two sufficiently off resonant lasers, is substantially not involved in the system dynamics. The more the lasers are off resonant, the more the non populated level may be considered as a \emph{virtual} one, in the sense that no real transition to it occurs.
Consider now the Heisenberg equation of motion for the transition operator $\hat{\sigma}_{-v}^{(H)}$:
\begin{equation}\label{temp-der}
  i\frac{d \hat{\sigma}_{-v}^{(H)}}{dt}=(\omega_v-\omega_1)\hat{\sigma}_{-v}^{(H)}
       -g_a^* e^{i(\vett{k}_a\cdot\vett{r}-\omega_a t)}
           (\hat{\sigma}_{--}^{(H)}-\hat{\sigma}_{vv}^{(H)})
       -g_b^* e^{i(\vett{k}_b\cdot\vett{r}-\omega_b t)}\hat{\sigma}_{-+}^{(H)}
\end{equation}
and introduce the modified operators 
${\hat{\sigma}}_{-v}^{(R)}=e^{i\omega_{a}t}{\hat{\sigma}}_{-v}^{(H)}$ and ${\hat{\sigma}}_{+v}^{(R)}=e^{i\omega_{b}t}{\hat{\sigma}}_{+v}^{(H)}$, by which the explicit time dependence in the equation of motion may be removed.
In the light of the previous remarks we are legitimate to assume that the modified transition operators, relative to the level $\Ket{v}$, have temporal derivatives equal to zero. Writing down the equation of motion for ${\hat{\sigma}}_{-v}^{(R)}$
\begin{equation}\label{temp-der-rot}
  i\frac{d \hat{{\sigma}}_{-v}^{(R)}}{dt}=
       (\Delta_{-v})\hat{{\sigma}}_{-v}^{(R)}
       -g_a^* e^{i\vett{k}_a\cdot\vett{r}}
           (\hat{\sigma}_{--}^{(R)}-\hat{\sigma}_{vv}^{(R)})
       -g_b^* e^{i\vett{k}_b\cdot\vett{r}}\hat{{\sigma}}_{-+}^{(R)}
\end{equation}
we take advantage of this condition imposing $i\frac{d \hat{\tilde{\sigma}}_{-v}^{(R)}}{dt}=0$ due to the off resonance of the lasers. In this way we easily deduce

\begin{equation}
  \hat{\sigma}_{-v}^{(H)}=\frac{1}{\Delta_{-v}}\left[ 
      g_a^* e^{i(\vett{k}_a\cdot\vett{r}-\omega_a t)}
        (\hat{\sigma}_{--}^{(H)}-\hat{\sigma}_{vv}^{(H)})+
      g_b^* e^{i(\vett{k}_b\cdot\vett{r}-\omega_b t)}\hat{\sigma}_{-+}^{(H)}
                                               \right]
\end{equation}

An analogous expression may be reached for $\hat{\sigma}_{+v}^{(H)}$.
Eliminating, via such relations, the $\Ket{\pm}\longleftrightarrow\Ket{v}$ transition operators in the Heisenberg picture expression of the Hamiltonian model and coming back to the Schr\"odinger picture, one finally obtains
\begin{eqnarray}\label{Ham-Effective}
  \hat{H}=\overline{H}_0+\hbar\tilde{\omega}_v\KetBra{v}{v}+\hat{H}_I^{(s)}
\end{eqnarray}
with
\begin{eqnarray}
  \overline{H}_0=\hbar{\tilde{\omega}}_1\KetBra{-}{-}+\hbar{\tilde{\omega}}_2\KetBra{+}{+}             
             +\hbar\omega_0\NumSum{a}
\end{eqnarray}
\begin{eqnarray}
  \left\{
  \begin{array}{l}
  \tilde{\omega}_{1}=\omega_{1}-\frac{2|g_a|^2}{\Delta_{-v}}
                 \;\;\;
  \left(\tilde{\omega}_{2}=\omega_{2}-  \frac{2|g_b|^2}{\Delta_{+v}}\right)\\
  \tilde{\omega}_v=\omega_v+\frac{|g_a|^2}{\Delta_{-v}}+\frac{|g_b|^2}{\Delta_{+v}}\\
  g=g_a g_b^* \left( \frac{1}{\Delta_{-v}}+ \frac{1}{\Delta_{+v}} \right)
  \end{array}
  \right.
\end{eqnarray}
\begin{eqnarray}\label{ham-int}
  \hat{H}_I^{(s)} = -\left[\hbar g e^{-i(\vett{k}_L\cdot \vett{r}-\omega_{L}t)}\KetBra{-}{+}
         + h.c. \right]
\end{eqnarray}
\begin{eqnarray}
  \label{effparam}
  \left\{
  \begin{array}{l}
  \vett{k}_{L}=\vett{k}_{a}-\vett{k}_{b}\\
  \omega_{L}=\tilde{\omega}_{a}-\tilde{\omega}_{b}
  \end{array}
  \right.
\end{eqnarray}

The effective coupling reached in eq.(\ref{ham-int}) formally describes a dipolar coupling interaction between the atomic levels $\Ket{+}$ and $\Ket{-}$ due to a \emph{fictitious} laser of parameters given by eq.(\ref{effparam}).
It is worth noting that the adiabatic elimination introduces the so called Raman Stark shifts, well visible in the modification of the atomic level eigenvalues $\hbar\tilde{\omega}_j$, $j=1,2,v$.

Exploiting the Baker-Hausdorff-Campbell lemma the exponential in eq.(\ref{ham-int}) becomes
\begin{eqnarray}
  \nonumber
  e^{-i\vett{k}_L\cdot\vett{r}}&=&e^{-i\eta_L(\hat{a}_L+\hat{a}_L^{\dag})}=
  e^{-\frac{\eta_L^2}{2}}e^{-i\eta_L\hat{a}_L}e^{-i\eta_L\hat{a}_L^{\dag}}=\\
  &=&e^{-\frac{\eta^2}{2}}\sum_{s,j=0}^{\infty}
  \frac{(-i\eta_L)^{s+j}}{s!j!}(\hat{a}_L^\dag)^j\hat{a}_L^s
\end{eqnarray}
with $\eta_L=\sqrt{\frac{\hbar}{2 M \omega_0}}|\vett{k}_L|$ the Lamb-Dicke parameter, $M$ the mass of the ion, and where the creation (annihilation) operators, 
$\hat{a}_L^{\dag}=cos(\theta)\hat{a}_z^{\dag} +
 sin(\theta)cos(\phi)\hat{a}_x^{\dag}         + sin(\theta)sin(\phi)\hat{a}_y^{\dag}$
($\hat{a}_L$), relative to the vibrational motion in the direction $L$ (that of $\vett{k}_L$) have been introduced, $\theta$ and $\phi$ being the $\vett{k}_L$ vector azimuthal and polar angles respectively. 
Here we also define the number operator $\hat{n}_L=\hat{a}_L^\dag\hat{a}_L$.
Passing to the interaction picture relative to $\tilde{H}_0$, we have

\begin{eqnarray}\label{ham-total}
  \nonumber
  \hat{H}_I^{(s)}\rightarrow
  \hat{H}_I =\\ 
    -\hbar g e^{-\frac{\eta_L^2}{2}}
    \sum_{s,j=0}^{\infty}\frac{(-i\eta_L)^{s+j}}{s!j!}
    (\hat{a}_L^\dag)^j\hat{a}_L^s
    e^{i(j-s)\omega_0 t}
    \KetBra{-}{+}e^{-i(\tilde{\omega}_A-\omega_L)t} + h.c.
\end{eqnarray}
where the free time evolutions of the creation (annihilation) operators and atomic transition operators have been introduced, and $\tilde{\omega}_A=\tilde{\omega}_b-\tilde{\omega}_a$.

It is well known\cite{VogelModelli,Knight} that, assuming $g << \omega_0$ and choosing the laser frequencies in such a way that $\omega_L=\tilde{\omega}_A$, one finally obtains
\begin{eqnarray}\label{ham-total-2}
  \hat{H}_I = 
    -\hbar g e^{-\frac{\eta_L^2}{2}}
    \sum_{j=0}^{\infty}\frac{(i\eta_L)^{2j}}{j!^2}
    (\hat{a}_L^\dag)^j\hat{a}_L^j
    (\KetBra{-}{+}+\KetBra{+}{-})
\end{eqnarray}
where the rapidly oscillating terms have been discarded. Observe that neglecting in eq.(\ref{ham-total}) the $j\not=s$ terms is,
\emph{from a mere technical point of view}, an approximation similar to the standard RWA, just in the sense that in both cases rapidly oscillating terms are discarded.
Many experimental results\cite{nist, nist-exp} validate indeed the realibility of this largely used approximation.

Summarising we have used a Raman scheme to effectively couple two levels, between which dipole transitions are forbidden, via a third auxiliary level. Moreover, suitably tuning the lasers, we have obtained a Hamiltonian model responsible for inducing vibrational energy dependent atomic transitions.

It is important to observe that, considering eq.(\ref{effparam}), the laser describing the effective coupling between $\Ket{-}$ and $\Ket{+}$ has frequency and wave vector which may be changed independently. Such a circumstance and the additivity of the interaction energy terms both imply the possibility of implementing a lot of vibronic coupling of the form $\hat{H}_I=f(\hat{n}_L)\hat{\sigma}_x$, with $\hat{\sigma}_x=\KetBra{-}{+}+\KetBra{+}{-}$.

\section{Implementing specular reflection operators}

The Hamiltonian model

\begin{eqnarray}\label{Ham-Parity}
  \hat{H}_I = \hbar g\hat{n}_L\hat{\sigma}_x
\end{eqnarray}
generates the temporal evolution operator
\begin{eqnarray}
  \hat{U}(t) = e^{-i g\hat{n}_L\hat{\sigma}_x}
\end{eqnarray}
responsible for a dynamics useful in order to implement a specular reflection operator. Indeed suppose the system prepared in the state
\begin{equation}
  \Ket{\psi(t=0)}=\sum_{\vett{n}} c_{\vett{n}}\Ket{\vett{n}} \Ket{-}
\end{equation}
where $\Ket{\vett{n}}\equiv\Ket{n_x,n_y,n_z}$ is a Fock state with $n_i$ bosonic excitations along $i=x,y,z$. Let the Hamiltonian operator acts on the vibrational quanta (phonons) relative to the oscillations along $z$. Under these assumptions one has

\begin{eqnarray}\label{EvTemp-CombLin}
  \nonumber
  \hat{U}(t)\Ket{\psi(t=0)}&=&
                   \left[\sum_{\vett{n}}
        c_{\vett{n}}\cos(\gamma n_z t)\Ket{\vett{n}}
                    \right]
                    \Ket{-}\\
                &-&i \left[\sum_{\vett{n}}
        c_{\vett{n}}\sin(\gamma n_z t)\Ket{\vett{n}}
                   \right]
                     \Ket{+}
\end{eqnarray}

After an interaction time $t=\frac{\pi}{g}$ it results
\begin{eqnarray}\label{EvTemp-CombLin-pi2}
  \Ket{\psi\left(t=\frac{\pi}{g}\right)}=
            \hat{U}\left(t=\frac{\pi}{g}\right)\Ket{\psi(t=0)}=
                   \left[\sum_{\vett{n}}
        c_{\vett{n}}(-1)^{n_z}\Ket{\vett{n}}
                    \right]
                    \Ket{-}
\end{eqnarray}

It is easy to convince oneself that this is just the action of a specular reflection (axis inversion) with respect to the plane $xy$ (the axis $z$). Indeed the eigenfunctions of a harmonic oscillator are well defined parity states having the same parity of the number of phonons.
In the light of these results, in order to realize a specular reflection operator, our target is now to implement the time evolution described by $\hat{U}(t)$ for a time interval $t=\frac{\pi}{g}$. To this purpose consider the action of two Raman schemes characterized by two effective strengths $g_1$ and $g_2$, frequencies both equal to the Raman Stark shifted atomic frequency and wave vectors both directed along $z$ but having different moduli. The resulting Hamiltonian model in the interaction picture is
\begin{eqnarray}\label{ham-total-3}
  \nonumber
  \hat{H}_I = 
    &-&\hbar g_1 e^{-\frac{\eta_1^2}{2}}
    \sum_{j=0}^{\infty}\frac{(i\eta_1)^{2j}}{j!^2}
    (\hat{a}_z^\dag)^j\hat{a}_z^j
    (\KetBra{-}{+}+\KetBra{+}{-})\\
    &-&\hbar g_2 e^{-\frac{\eta_2^2}{2}}
    \sum_{j=0}^{\infty}\frac{(i\eta_2)^{2j}}{j!^2}
    (\hat{a}_z^\dag)^j\hat{a}_z^j
    (\KetBra{-}{+}+\KetBra{+}{-}) 
\end{eqnarray}
where $\eta_i$ (with $i=1,2$) is the Lamb-Dicke parameter relative to the wave vector $\vett{k}_i$ of the $i$-th effective laser.  

At this point we can set the coupling strengths satisfying the condition
$g_1{\left(1-\frac{\eta_1^2}{2} \right)}+g_2{\left(1-\frac{\eta_2^2}{2} \right)}=0$. Assuming the Lamb-Dicke limit, meaning that $\eta_i<<1$, and neglecting the terms of the two series smaller than $\eta_i^2$, one obtains the interaction hamiltonian in eq.(\ref{Ham-Parity}) with $g=-g_1(\eta_2^2-\eta_1^2)$.
In this way we have implemented the Hamiltonian whose time evolution operator simulates the action of a specular reflection operator.

\section{Applications and Conclusive remarks}

The proposal presented in the previous section has an intrinsic importance and moreover furnishes the starting point of some useful applications in quantum state manipulation. Indeed it permits to orthogonalize a class of quantum states and may be also used to realize a temporal reversion process with respect to a huge class of time evolutions.


In the last years many efforts have been done in order to realize applications in quantum logic\cite{nist,logic}. In such a context the study of the dynamics of a qubit, which is nothing but a two level system, plays a fundamental role since it permits to realize quantum logic gates as for example the NOT-gate, whose effect on a qubit is to \emph{invert} it, in the sense that it returns as output the state orthogonal to the input state\cite{U-Not}. 

An important application of the Parity operator implementation just concernes the realization of a NOT-gate.
Consider a qubit defined by the two possible \lq\lq values\rq\rq (meaning states)
\begin{equation}
  \Ket{\psi_{\pm}}=\frac{1}{\sqrt{2}}\left(\Ket{\phi_e}\pm\Ket{\phi_o}\right)
\end{equation}
where $\Ket{\phi_e}$ and $\Ket{\phi_o}$ are appropriately prefixed normalized vibrational states of the trapped ion center of mass such that $\hat{\Pi}_L\Ket{\phi_e}=\Ket{\phi_e}$ and $\hat{\Pi}_L\Ket{\phi_o}=-\Ket{\phi_o}$. Thus $\BraKet{\phi_e}{\phi_o}=0$ and hence $\BraKet{\psi_+}{\psi_-}=0$

The space reversal operator, $\hat{\Pi}_L$, transforms the state $\Ket{\psi_{+}}$ into $\Ket{\psi_{-}}$ and vice versa and thus its implementation, as here reported, allows to effectively \emph{invert} the input state as a NOT-gate does.


Another application concerns the possibility of using $\hat{\Pi}_L$ to restore an initial condition after the system has left it. In addition to the intrinsic theoretical importance of such a possibility, reversing a temporal evolution may turn to be useful in quantum state manipulation.

Consider a time evolution operator $T(t)=e^{-\frac{i}{\hbar}\hat{H}t}$ ($\hat{H}$ being the time independent Hamiltonian of the system) acting upon a general state $\Ket{\psi_0}$, so that $\Ket{\psi(t)}=T(t)\Ket{\psi_0}$ is the state of the trapped ion at the time $t$.
Suppose now that $\hat{\Pi}_L$ anticommute with the generator of $T(t)$ (that is, the Hamiltonian). In such a situation and recalling that $\hat{\Pi}_L^{\dag}=\hat{\Pi}_L$, it immediately follows that
\begin{equation}
  \hat{\Pi}_L T(t) \hat{\Pi}_L=T^{-1}(t)
\end{equation}

Hence the time reversal operator may be practically realized via the subsequent application of three pulses on the system in the state $\Ket{\psi(t)}$: a specular reflection operator, a time evolution operator identical to the one we want to reverse and again a specular reflection.
Although in many cases the temporal reversal may be realized following other procedures, we emphasise that our method allows to reverse a huge class of time evolutions without substantially changing the experimental apparatus.


Throughout this paper we have assumed the possibility of treating the system as an isolated and decoherence-free one. Such an assumption has to be verified in order to claim that our proposal may be of experimental interest.
Taking $\eta_1\approx 0.2$, $\eta_2\approx 0.3$ and $\hbar g_1\approx \hbar g_2\approx 3 MHz$ from ref\cite{nist-exp}, where the authors deal with a $Be^+$ trapped ion, we obtain $\hbar g\approx 150 KHz$ and an interaction time of about $20\mu s$, during which the system exhibits coherent behavior as clearly shown by the experimental results in the just recalled reference.
From ref.\cite{nist-exp} we also take typical values for $\omega_0\approx 11.2 MHz$, $\omega_A\approx 1.2 GHz$, and for detunings, $\Delta\approx 12 GHz$. The value of the trap parameter $\omega_0$ satisfies the condition $g << \omega_0$ under which eq.(\ref{ham-total-2}) has been deduced.
It is worth noting that the assumed values of the Lamb-Dicke parameters $\eta_1$ and $\eta_2$ are also compatible with the Lamb-Dicke limit approximation. 
This is due to the fact that truncating the series in eq.(\ref{ham-total-2}) requires $\eta_i^2<<1$ which is certainly true in the experimental conditions under scrutiny.
Concluding, we believe our experimental scheme for the realization of specular reflection operators has a practical usefulness and theoretical importance and, moreover, an acceptable degree of feasibility.

\section*{Acknowledgements}
We thank S.Maniscalco for stimulating discussions.
One of the authors (A. N.) acknowledges financial support from Finanziamento Progetto Giovani Ricercatori anno 1998, Comitato 02.

\end{document}